\RequirePackage{lineno}
\documentclass[twocolumn,aps,prc,showpacs,superscriptaddress,floatfix]{revtex4-2}
\usepackage{rotating}
\usepackage{epsfig}
\usepackage[colorlinks=true, allcolors=blue]{hyperref}
\usepackage{lineno}
\usepackage[normalem]{ulem}

\usepackage{amsmath}
\usepackage{graphicx}
\usepackage[caption=false]{subfig}
\graphicspath{ {./}, {./image/NotePlots} }

\usepackage{xcolor}

\usepackage{xparse,xcoffins}

\ExplSyntaxOn
\NewCoffin\imagecoffin
\NewCoffin\labelcoffin

\keys_define:nn { fqwang/label }
 {
  label   .tl_set:N = \l_fqwang_label_tl,
  labelbox .bool_set:N = \l_fqwang_label_box_bool,
  labelbox .default:n = true,
  fontsize .tl_set:N = \l_fqwang_label_size_tl,
  fontsize .initial:n = \normalsize,
  pos .choice:,
  pos/nw .code:n = \tl_set:Nn \l_fqwang_label_pos_tl { left,up },
  pos/ne .code:n = \tl_set:Nn \l_fqwang_label_pos_tl { right,up },
  pos/sw .code:n = \tl_set:Nn \l_fqwang_label_pos_tl { left,down },
  pos/se .code:n = \tl_set:Nn \l_fqwang_label_pos_tl { right,down },
  pos/nwhigh .code:n = \tl_set:Nn \l_fqwang_label_pos_tl { left,uphigh },
  pos/nehigh .code:n = \tl_set:Nn \l_fqwang_label_pos_tl { right,uphigh },
  pos/swhigh .code:n = \tl_set:Nn \l_fqwang_label_pos_tl { left,downhigh },
  pos/sehigh .code:n = \tl_set:Nn \l_fqwang_label_pos_tl { right,downhigh },
  pos/nwlow .code:n = \tl_set:Nn \l_fqwang_label_pos_tl { left,uplow },
  pos/nelow .code:n = \tl_set:Nn \l_fqwang_label_pos_tl { right,uplow },
  pos/swlow .code:n = \tl_set:Nn \l_fqwang_label_pos_tl { left,downlow },
  pos/selow .code:n = \tl_set:Nn \l_fqwang_label_pos_tl { right,downlow },
  pos/nwHigh .code:n = \tl_set:Nn \l_fqwang_label_pos_tl { left,upHigh },
  pos/neHigh .code:n = \tl_set:Nn \l_fqwang_label_pos_tl { right,upHigh },
  pos/swHigh .code:n = \tl_set:Nn \l_fqwang_label_pos_tl { left,downHigh },
  pos/seHigh .code:n = \tl_set:Nn \l_fqwang_label_pos_tl { right,downHigh },
  pos/nwLow .code:n = \tl_set:Nn \l_fqwang_label_pos_tl { left,upLow },
  pos/neLow .code:n = \tl_set:Nn \l_fqwang_label_pos_tl { right,upLow },
  pos/swLow .code:n = \tl_set:Nn \l_fqwang_label_pos_tl { left,downLow },
  pos/seLow .code:n = \tl_set:Nn \l_fqwang_label_pos_tl { right,downLow },
  pos/n .code:n = \tl_set:Nn \l_fqwang_label_pos_tl { hc,up },
  pos/w .code:n = \tl_set:Nn \l_fqwang_label_pos_tl { left,vc },
  pos/s .code:n = \tl_set:Nn \l_fqwang_label_pos_tl { hc,down },
  pos/e .code:n = \tl_set:Nn \l_fqwang_label_pos_tl { right,vc },
  pos .initial:n = nw,
  unknown .code:n   = \clist_put_right:Nx \l_fqwang_label_clist
                       { \l_keys_key_tl = \exp_not:n { #1 } }
 }

\clist_new:N \l_fqwang_label_clist
\box_new:N \l_fqwang_label_box
\box_new:N \l_fqwang_label_image_box

\NewDocumentCommand{\xincludegraphics}{O{}m}
 {
  \group_begin:
  \tl_clear:N \l_fqwang_label_tl
  \clist_clear:N \l_fqwang_label_clist
  \keys_set:nn { fqwang/label } { #1 }
  \tl_if_empty:NTF \l_fqwang_label_tl
   {
    \fqwang_includegraphics:Vn \l_fqwang_label_clist { #2 }
   }
   {
    \SetHorizontalCoffin\imagecoffin
     {
      \fqwang_includegraphics:Vn \l_fqwang_label_clist { #2 }
     }
    \SetHorizontalCoffin\labelcoffin
     {
      \raisebox{\depth}
       {
        \bool_if:NTF \l_fqwang_label_box_bool
         { \fcolorbox{white}{white}{\l_fqwang_label_size_tl\l_fqwang_label_tl} }
         { \l_fqwang_label_size_tl\l_fqwang_label_tl }
       }
     }
    \SetVerticalPole\imagecoffin{left}{36pt+\CoffinWidth\labelcoffin/2}
    \SetVerticalPole\imagecoffin{right}{\Width-36pt-\CoffinWidth\labelcoffin/2}
    \SetHorizontalPole\imagecoffin{up}{\Height-12pt-\CoffinHeight\labelcoffin/2}
    \SetHorizontalPole\imagecoffin{down}{12pt+\CoffinHeight\labelcoffin/2}
    \SetHorizontalPole\imagecoffin{uphigh}{\Height-6pt-\CoffinHeight\labelcoffin/2}
    \SetHorizontalPole\imagecoffin{downlow}{6pt+\CoffinHeight\labelcoffin/2}
    \SetHorizontalPole\imagecoffin{uplow}{\Height-18pt-\CoffinHeight\labelcoffin/2}
    \SetHorizontalPole\imagecoffin{downhigh}{12pt+\CoffinHeight\labelcoffin/2}
    \SetHorizontalPole\imagecoffin{upHigh}{\Height-0pt-\CoffinHeight\labelcoffin/2}
    \SetHorizontalPole\imagecoffin{downLow}{0pt+\CoffinHeight\labelcoffin/2}
    \SetHorizontalPole\imagecoffin{upLow}{\Height-24pt-\CoffinHeight\labelcoffin/2}
    \SetHorizontalPole\imagecoffin{downHigh}{42pt+\CoffinHeight\labelcoffin/2}
    \use:x{\JoinCoffins\imagecoffin[\l_fqwang_label_pos_tl]\labelcoffin[vc,hc]} 
    \TypesetCoffin\imagecoffin
   }
   \group_end:
 }
\NewDocumentCommand{\setlabel}{m}
 {
  \keys_set:nn { fqwang/label } { #1 }
 }
\cs_new_protected:Nn \fqwang_includegraphics:nn
 {
  \includegraphics[#1]{#2}
 }
\cs_generate_variant:Nn \fqwang_includegraphics:nn { V }

\ExplSyntaxOff

\newcommand\mean[1] {\left\langle#1\right\rangle}

\newcommand\snn     {\sqrt{s_{_{\rm NN}}}}
\newcommand\gevc    {GeV/$c$}
\newcommand\gevcc   {GeV/$c^2$}
\newcommand\minv    {m_{\rm inv}}
\newcommand\pt      {\ensuremath{p_{\perp}}}
\newcommand\ptpair  {p_{\perp,{\rm pair}}}
\newcommand\bkg     {{\rm bkg}}
\newcommand\kk      {{K^+K^-}}
\newcommand\df      {Data Folding}
\newcommand\ds      {Data Scaling}

\newcommand\rzero   {\rho_{00}}
\newcommand{\Drho} {\Delta\rho}

\begin{document}

\title{Exploring Data-Driven Corrections for $\phi$-Meson Global Spin Alignment Measurements}

\author{C.W.~Robertson}
\email{rober558@purdue.edu}
\affiliation{Department of Physics and Astronomy, Purdue University, West Lafayette, IN 47907}
\author{Yicheng Feng}
\email{feng216@purdue.edu}
\affiliation{Department of Physics and Astronomy, Purdue University, West Lafayette, IN 47907}
\author{Fuqiang Wang}
\email{fqwang@purdue.edu}
\affiliation{Department of Physics and Astronomy, Purdue University, West Lafayette, IN 47907}

\begin{abstract}
Non-central heavy ion collisions generate large orbital angular momentum (OAM), providing opportunities to study spin phenomena such as the global spin alignment of vector mesons. Such studies are expected to reveal properties of the quark-gluon plasma produced in these collisions. 
Global spin alignment of vector mesons, such as the $\phi$-meson, can be measured by the $00^{\rm th}$ coefficient of the spin density matrix,  $\rzero$, via the polar angle of the decay kaon momentum in the parent rest frame with respect to the OAM direction of the collision. A deviation of $\rzero$ from the isotropic value of $1/3$ indicates a finite spin alignment. The reported signal of $\rzero-1/3$ is on the order of $\sim 1\%$ and therefore corrections for finite detector performance and acceptance, which are expected to be on the order of a few tenths of a percent, are important. Additional complications in the detector corrections may arise from the $\phi$-meson azimuthal anisotropy which could become intertwined with the detector efficiency. Typically, detector corrections for global spin alignment of vector mesons are performed with {\em Monte-Carlo} (MC) methods using detector simulation packages such as {\sc geant}, however it is unclear if such methods can be trusted at the needed level of precision. In this paper, we investigate an alternative, data-driven approach in correcting for detector effects. This approach utilizes detector effects on combinatorial kaon pairs from $\phi$-meson decays that fall within the $\phi$-meson mass window, which can be obtained through statistical identification of decay kaons in real data analysis. We examine the degree of success of such a data-driven approach using toy-model MC simulations as well as its shortcomings.
\end{abstract}
%\pacs{25.75.-q,25.75.Ld}% PACS

\maketitle

%========================================================================================

\section{Introduction}
Relativistic heavy ion collisions create a hot and dense medium where quarks and gluons are ``freed'' over an extended volume comparable to the size of heavy nuclei~\cite{Arsene:2004fa,Back:2004je,Adams:2005dq,Adcox:2004mh,Muller:2012zq,Roland:2014jsa}. Such a state of matter, called the quark-gluon plasma (QGP), is believed to permeate the early universe after the Big Bang for a period of $\sim$10~$\mu$sec, at which the primordial matter hadronized into particles like the protons and neutrons we know today. Studies of the QGP created in heavy ion collisions promise to reveal fundamental properties of quantum chromodynamics (QCD), the theory known to govern the interactions of quarks and gluons.

In non-central heavy ion collisions, a large orbital angular momentum (OAM) is present~\cite{Liang:2004ph,Voloshin:2004ha}. 
The vorticity field generated by the large OAM in the created QGP can polarize the spin-1/2 quarks~\cite{Wu:2019eyi,Sheng:2019kmk,Becattini:2020ngo}. 
This polarization is inherited by final-state hadrons via hadronization, the effect of which can be measured by parity-violating weak decays of hyperons and by parity-conserving strong decays of vector mesons~\cite{Liang:2004ph}. 
A finite global spin polarization of the $\Lambda$-hyperon has  indeed been observed, on the order of 1\%, suggesting the presence of an ultra-strong vortical field in the QGP~\cite{STAR:2017ckg}. 
A finite global spin alignment of the $\phi$-meson has been recently reported by the ALICE~\cite{ALICE:2019aid} and STAR~\cite{STAR:2022fan}  experiments. 
Spin alignment of vector mesons is a result of spin-spin correlations, which are naively expected to be on the order of the square of the spin polarization~\cite{Sheng:2019kmk}. 
The reported $\phi$-meson spin alignment is, however, on the order of 1\%, much larger than the expected value of $10^{-4}$ from the square of the observed $\Lambda$-hyperon spin polarization. This prompted the suggestion of strong color field fluctuations as a plausible novel physics mechanism for the large spin alignment~\cite{Sheng:2019kmk,Sheng:2022wsy}.

The global spin alignment is measured by the angular distribution of a decay kaon from the $\phi\rightarrow\kk$ decay~\cite{Liang:2004ph,ALICE:2019aid,STAR:2022fan},
\begin{equation}
    \frac{dN}{d\cos{\theta^{*}}} \propto (1-\rho_{00}) + (3\rho_{00}-1)\cos^{2}{\theta^{*}}\,,
	\label{eq:dCosThe}
\end{equation}
where $\theta^*$ is the polar angle of the  kaon's momentum vector in the $\phi$-meson rest frame with respect to the global OAM in the lab frame. 
The parameter $\rho_{00}$ is the $00^{\rm th}$ coefficient of the spin density matrix. A uniform angular distribution gives $\rho_{00}=1/3$. Deviation of $\rho_{00}$ from 1/3 indicates a finite spin alignment; we use $\Drho\equiv\rho_{00} -1/3$ to quantify the deviation.
The OAM direction is on average perpendicular to the reaction plane (RP)~\cite{Liang:2004ph,Voloshin:2004ha}, the plane spanned by the impact parameter direction of the collision and the beam axis. 
Since the RP is not measured, the first-order harmonic plane measured by neutron deposition in the zero-degree calorimeters is often used as a proxy for the RP~\cite{Poskanzer:1998yz}. 
The second-order harmonic plane corresponding to elliptic flow of produced particles is also often used~\cite{Ollitrault:1993ba,Poskanzer:1998yz}.
The global spin alignment is thus measured with respect to the normal of the first-order or second-order harmonic plane~\cite{ALICE:2019aid,STAR:2022fan}.

Conventionally, $\phi$-meson yield is measured in bins of $\cos{\theta^{*}}$, corrected by $\phi$-meson reconstruction efficiency and detector acceptance~\cite{ALICE:2019aid,STAR:2022fan}. These detector effects can be obtained from an embedding technique--embedding {\em Monte Carlo} (MC) generated $\phi$ mesons into real data, simulating the detector responses, and reconstructing some of the $\phi$ mesons from those decay kaons that survive the detector. The correction to the $\phi$ meson yield in each $\cos{\theta^{*}}$ bin is then applied.

In experiment, the product of the single kaon efficiencies, properly weighted according to the $\phi$-decay kinematics, is often used as a substitute for the $\phi$-meson efficiency because of a lack of $\phi$-meson embedding statistics. This was done in the STAR publication of the $\phi$ spin alignment~\cite{STAR:2022fan}. 
In general, the $\phi$-meson embedding efficiency (effectively two-particle efficiency) can be fairly well reproduced by the convoluted product of two single-kaon efficiencies. Despite this agreement, it is unclear how well the two-particle effects have been modeled in embedding MC because those effects may be missing in both $\phi$ embedding and single kaon embedding.
Since the kaon efficiency is obtained for a given acceptance cut such as $|\eta|<1$, those outside the acceptance cannot be corrected back by the kaon efficiency. This is not the case for the $\phi$-meson embedding efficiency where the $\phi$ mesons whose decay daughters fall out of the acceptance are properly accounted for. Such an acceptance effect using kaon embedding efficiencies needs to be corrected by other means, and STAR used the Pythia simulation, properly weighted by measured $\phi$-meson kinematics, to obtain the acceptance effect~\cite{STAR:2022fan}.

Embedding efficiencies have been used extensively for single-particle spectra measurements~\cite{STAR:2003jwm,STAR:2008med}. 
The embedding technique simulates detector hit, track merging, and track splitting information from MC tracks. 
It is fairly reliable once the MC is tuned to faithfully reproduce data measurements of tracking properties, such as the distributions of the number of hits used for track reconstruction and the distance of closest approach of tracks from the primary vertex~\cite{STAR:2008med}. 

This is easy to do on single-particle level, however, it is not clear whether embedding can adequately describe two-particle distributions which are less thoroughly examined. 
For example, close tracks can be misidentified as a single track, an effect referred to as track merging; likewise, a single track can be reconstructed as two separate tracks, an effect referred to as track splitting~\cite{STAR:2001gzb}. 
These track merging and splitting effects may not be adequately simulated in embedding MC. 

Such an inadequacy may not be important to single-particle spectra measurements such as the $\phi$-meson's as the angular space where such two-particle effects become important may be negligible to single-particle yield measurement. 
However, it is non-trivial to assess the effects on correlation measurements such as the $\phi$ spin alignment. 
This is particularly important for $\phi$ spin alignment measurement, for a number of reasons:
\begin{itemize}
    \item $\theta^*$ is a 3-D polar angle in the $\phi$-meson rest frame, involving boost from the laboratory frame where the tracks are measured and characterized. The boost is in turn determined by the measured track parameters. The relationship between $\theta^*$ and the measured kaon tracks is complex and it is opaque how  imperfections in tracking and acceptance propagate to the $\theta^*$ measurements.
    \item Since the $\Drho$ signal is small, on the order of 1\% as reported~\cite{STAR:2022fan}, it is unclear if the standard embedding can be trusted on that level for 3-D kinematics. 
    
    \item Because the detector correction is now applied as a function of $\cos{\theta^{*}}$, the efficiency and acceptance effects are likely dependent of the azimuthal anisotropy ($v_2$) of the $\phi$ mesons, beyond the trivial detector occupancy effect arising from the anisotropic azimuthal distributions of particles. The efficiency and acceptance effects may also depend on the $\rho_{00}$ signal strength because it alters the angular distributions of the decay kaons.
\end{itemize}

In principle, detector effects obtained from embedding MC cannot be rigorously examined against data because the data {\em truth} is unknown.
Detector effects on single-particle measurements may be fairly reliably modeled by embedding MC, and the systematic uncertainties of these effects can be faithfully assessed. 
For example, there are systematic uncertainties due to subtle mismatches between MC and data. Additionally, effects such as TPC gas mixture, pressure, and humidity fluctuations are assessed with a typical systematic uncertainty on the order of 5\%~\cite{STAR:2008med}. 
However, as aforementioned, two-particle detector effects on correlation measurements are less thoroughly studied. 
It would be desirable to have a data-driven way to correct for detector effects, which  in principle has all the single- and multi-particle detector effects built in by real data. Data-driven correction methods, if successful, are always better and thus preferred than any simulation methods because all real-world effects are by definition included in real data.
In this article, we explore a data-driven method of correcting for detector effects in $\phi$-meson global spin alignment measurements.

%=========================================================================================
\section{Invariant Mass Method \label{sec:InvMassMethod}}

In the conventional method, detector corrections are applied on to the $\phi$-meson yield as a function of $\cos\theta^*$, and then the $\rzero$ is extracted~\cite{ALICE:2019aid,STAR:2022fan}. It is a two-step process, and therefore it is probably difficult to apply a data-driven correction. The sole interest in $\phi$-meson spin alignment measurement is the quantity $\rho_{00}$. The idea of a data-driven correction is to apply a correction factor directly on to this quantity.
Instead of the two-step conventional method, such a quantity can be obtained by calculating the average $\mean{\cos^2{\theta^{*}}}$ as a function of the invariant mass of $\kk$ pairs ($\minv$), which can be readily converted into $\rho_{00}$ via Eq.~(\ref{eq:dCosThe}), namely,
\begin{equation}
    \Drho \equiv \rzero - \frac{1}{3} = \frac{2}{5}\left( \mean{\cos ^2\theta^*} - \frac{1}{3} \right)\,.
    \label{eq:conv}
\end{equation}
The correction factor for detector effects is then applied directly on to the raw $\rzero$ value to obtain the final $\rzero$ measurement.
This is called the invariant mass method~\cite{Robertson:2025fxm}. 

The invariant mass method is described in Ref.~\cite{Robertson:2025fxm}. Briefly, the signal to background ratio of the $\phi$-meson, $r(\minv)=\mbox{Breit-Wigner}/f_\bkg$, is extracted by fitting the $\minv$ distribution with a Breit-Wigner signal atop a combinatorial background function $f_\bkg$. The profile of $\mean{\cos ^2\theta^*}$ or $\Drho$ vs $\minv$ is obtained.
The $\phi$-meson $\Drho$ can then be extracted by fitting the profile of $\Drho(\minv)$ vs $\minv$ to
\begin{equation}
    \Drho(\minv) = \frac{\Drho_{\bkg}(\minv) + r(\minv)\Drho}{1+r(\minv)}\,,
    \label{eq:InvMass}
\end{equation}
where $\Drho_{\bkg}(\minv)$ is the background shape of $\Drho$ as a function of $\minv$. The $\Drho_{\bkg}(\minv)$ functional form is unknown {\em a priori}, and assumptions (such as linear function) are made guided by data. 

In a perfect world, the invariant mass method and the yield method would give the same $\rzero$. In reality they differ because of different ways of handling backgrounds (effectively, different assumptions) and other sources of systematics, and thus provide good cross checks~\cite{Robertson:2025fxm}. 
In addition, the invariant mass method provides a convenient way for data-driven correction procedures.

%======================================================================================
\section{Data-Driven Correction Method \label{sec:DataDrivenMethod}}

The question we aim to address is: what is the change in $\Drho$ of the real $\phi$-meson decay $\kk$ pairs due to detector effects? Obviously, this question cannot be answered using kaon pairs from real $\phi$ decays in a data-driven way because the real $\phi$-meson $\Drho$ is unknown, and this is the exact physics signal one is trying to measure. We propose to use kaon pairs that are not from real $\phi$-meson signal but otherwise ``equal'' to those real signal kaon pairs. {\em The idea is to analyze the $\Drho$ of combinatorial pairs of kaons from $\phi$-meson decays measured in the detector within the $\phi$-meson mass region, and compare the result to that before the kaons suffer any detector effects.} We term these combinatorial kaon pairs ``pseudo-$\phi$ pairs,'' whereas ``real $\phi$ pairs'' refer to those real $\phi$-decay kaon pairs.
\begin{itemize}
    \item The former, i.e.~combinatorial pairs of kaons from $\phi$-meson decays measured in the detector, can be obtained from real data. There is one complication: the $\phi$-meson identification is statistical because of combinatorial background, so we cannot uniquely say which kaon is from a $\phi$-meson decay and which is not. To circumvent this, we use all identified kaons and scale them to match the decay kaon kinematics in $(\pt,\eta,\phi -\psi_2 )$ which is measured statistically. Here, $\psi_2$ is the second-order harmonic plane angle reconstructed in a real data analysis; for our purpose we simply use the generated $\psi_2$ (random from event to event) so we are free from resolution correction for the reconstructed event-plane (EP)~\cite{Poskanzer:1998yz} which is outside the scope of this study. We require these combinatorial pairs to be within the $\phi$ mass region (i.e.~pseudo-$\phi$) to ensure that their kinematics are as same as the real $\phi$-meson's in the $\phi$ rest frame. This procedure is referred to as ``\ds,'' and these pseudo $\phi$'s as ``data pseudo $\phi$'s.''
    \item The latter, i.e.~combinatorial pairs from $\phi$-decay kaons before any detector effects, can be obtained by MC sampling using published $\phi$-meson data. We require these combinatorial pairs to be within the $\phi$ mass region, i.e.~pseudo $\phi$'s.
    This procedure is referred to as  ``\df''. 
\end{itemize}
Once the $\Drho$ values are obtained, with and without detector effects, we can derive a correction simply from their difference.

There are two important {\em assumptions} in this data-driven correction procedure:
\begin{itemize}
    \item[{\em I)}] The detector effect on real $\phi$-decay pairs can be represented by that on the pseudo-$\phi$ kaon pairs. In other words, the pseudo-$\phi$ pairs are as similar to the real $\phi$ decay pairs as possible, the average detector effect seen by all pseudo-$\phi$ pairs, from a given sample of real $\phi$'s, would represent the detector effect on the real $\phi$. 
    \item[{\em II)}] The detector effect on pseudo-$\phi$ pairs can be obtained from data pseudo $\phi$'s, after proper weighting in kinematics. In other words, the combinatorial pairs of decay kaons are no different from those of all kaons regardless of their origins once the kinematics are matched.
\end{itemize}

In the following, we first describe the \df\ procedure in Section~\ref{sec:folding}, using a particular $\phi$-meson kinematic region for illustration. We then describe the \ds\ procedure in Section~\ref{sec:scaling}, in the way that a real data analysis would proceed, treating the generated MC data from \df\ as ``real'' data. 
By doing this, we are using identical input to \df\ and \ds.
In reality (a real data analysis), the input to \df\ is the published $\phi$-meson data accompanied by uncertainties, and the input to \ds\ is whatever the used data sample is. 
They are not identical, and the uncertainties of the published data need to be considered in assessing the uncertainties in the derived correction. This is, however, outside the scope of the present study, which focuses on the methodology of deriving data-driven corrections.

For reference, here we summarize the important terminology we use in this article:
\begin{itemize}
    \item {\em real $\phi$}: real $\phi$-mesons we generate with MC. See Section~\ref{sec:folding}.
    \item {\em pseudo $\phi$}: combinatorial $\kk$ pairs formed from decay kaons after rotating the $K^-$ by $\pi$ in azimuth (referred to as ``rotated pairs'') or from mixed events (``mixed pairs''), and within the $\phi$-meson mass window [1.015, 1.025]~\gevcc. See Section~\ref{sec:folding}. 

    \item {\em data pseudo $\phi$}: combinatorial $\kk$ pairs formed from all measured kaons in data after rotating the $K^-$ by $\pi$ in azimuth or from mixed events, and within the $\phi$-meson mass window [1.015, 1.025]~\gevcc, weighted to match the decay kaons in kinematics. 
    See Section~\ref{sec:scaling}.

    \item {\em survived real (pseudo) $\phi$}: real (pseudo) $\phi$'s that have survived detector effects. These are only accessible in our MC simulations and are used to assess the performance of our data-driven method. 
    See Section~\ref{sec:closure}. 
    Ideally, the detector effects on their $\Drho$ are equal ({\em Assumption I}) and equals to that on data pseudo $\phi$'s ({\em Assumption II}).
\end{itemize}

%---------------------------------------------------------------------------------------
\subsection{$\phi$-decay kaons without detector effects (\df)} \label{sec:folding}

To know the effects of imperfect detectors and finite acceptance, we need to know the {\em true} $\Drho$\ of these pseudo-$\phi$ kaon pairs before any detector effects. To do so, we generate $\phi$ mesons according to published data and decay them according to Eq.~(\ref{eq:dCosThe}) with a given $\rzero$. We then calculate the $\Drho$ of these pseudo-$\phi$ pairs as a function of $\minv$. %vs $\ptpair$. 
We use $\rzero=1/3$ (isotropic decays) for most of our studies; In Section~\ref{sec:DrhoDependence} we examine the effect of finite $\phi$-meson spin alignment on the detector correction procedure.

In practice, we need the $\phi$-meson $\pt$ spectrum (and $dN/dy$ multiplicity) and $v_2(\pt)$ in each centrality. For illustration, we use the $\phi$-meson data measured in 40-50\% centrality Au+Au collisions at $\snn=200$~GeV~\cite{STAR:2007mum}. We generate the $\phi$-meson $\pt$ from a fit to the measured $\pt$ spectrum. The STAR spin alignment measurements~\cite{STAR:2022fan} are confined within the $\phi$-meson $\pt$ range of $1.2 < \pt < 5.4$~\gevc, so we take this $\pt$ range for the purpose of generating $\phi$ mesons. The $\phi$-meson $v_2$ is taken to be simply proportional to $\pt$ with a cutoff at high $\pt$. For our purposes we use $v_2=0.064 \times \pt/$\gevc, where the coefficient (slope value) comes from a fit to $\phi$-meson 200 GeV data in the 10--40\% centrality measured in~\cite{STAR:2007mum} and we use constant $v_2$ for $\phi$-meson $\pt >3$~\gevc.
The average multiplicity density of $\phi$ mesons has been measured in~\cite{STAR:2008bgi}. We use  $dN_\phi/dy = 1.44$ corresponding to 40-50\% centrality 200 GeV Au+Au collisions. The event-by-event multiplicities within $|\eta|<1$ (two units of pseudorapidity) are generated from a Poisson distribution with twice this average.
For simplicity, the $\phi$ mesons are generated with a uniform pseudorapidity distribution within $|\eta|<1$.

We keep all decay kaons (no kinematic cuts on individual decay tracks) from $\phi$ mesons with kinematic selection described above. To get pseudo-$\phi$ kaon pairs, we rotate the $K^-$ by $\pi$ in azimuth and then use all $\kk$ pairs, referred to as ``rotated pairs.'' We could use all combinatorial kaon pairs (i.e., excluding the pair from a real $\phi$ decay) without rotating the $K^-$. We have verified that these two different ways give the same $\mean{\cos^2\theta^*}$ vs $\minv$ result. The primary reason to use the former is to be consistent with \ds\ where we only identify $\phi$ mesons statistically so cannot exclude the real $\phi$-decay kaon pairs; see Section~\ref{sec:scaling}. 
We boost the rotated kaon pair to its rest frame and calculate the $\mean{\cos^2\theta^*}$ vs $\minv$. We take the pseudo-$\phi$'s to be those rotated pairs within the pair mass window of $1.015<\minv<1.025$~\gevcc\ so that in each pair's rest frame the pseudo-$\phi$ kaon pair looks like a real $\phi$-decay pair.

We also use the mixed-event technique to form combinatorial kaon pairs, by taking a pair of $\kk$ from two different events. We refer to these pseudo-$\phi$ kaon pairs as ``mixed-event/mixed pairs.''
We follow the same procedure above for mixed events. 
There is an important distinction between mixed-event pairs and rotated pairs.
For mixed-event pairs, the EP is taken from the ``current'' event in calculating the $\mean{\cos^2\theta^*}$. 
Since the reaction planes are randomly generated, there is an EP mismatch between mixed events, which affects the corrections derived from mixed-event pseudo-$\phi$ pairs. 
One can of course use the RP angle in the generated events, but that is not practically useful because that cannot be done in real data analysis.
In real data analysis~\cite{STAR:2022fan}, the events can be rotated according to the reconstructed EP or they can be binned in the reconstructed EP angle before event mixing; however, because of the finite EP resolution, mixed events cannot be perfectly aligned with respect to the RP or the collision geometry. 
Generally, mixed events cannot possibly be made identical to real events in terms of {\em all} non-physics-related attributes of the events. 

In summary, the \df\ technique allows us to get the $\Drho$\ of pseudo-$\phi$ pairs without any detector effects of finite acceptance or track reconstruction inefficiency. 

%---------------------------------------------------------------------------------------
\subsection{$\phi$-decay kaons in ``real'' data (\ds)} \label{sec:scaling}
In this section, we describe how to calculate the $\Drho$ for pseudo-$\phi$ pairs in real data analysis. For illustration, in addition to the MC generated $\phi$ mesons, we also generate kaons to mimic the primordial kaons in real data. The primordial kaon $\pt$ distributions in 200 GeV Au+Au collisions have been measured previously and are well described by a Boltzmann distribution with a kinetic freeze-out temperature and a common radial flow velocity~\cite{STAR:2008med}. Here we generate primordial kaon $\pt$ from a Boltzmann distribution in the range $0< \pt < 10$~\gevc\ with values for the kinetic freeze-out temperature and collective boost velocity taken from 40-50\% centrality 200 GeV Au+Au measurements in Ref.~\cite{STAR:2008med}.
We use measurements in Ref.~\cite{STAR:2004jwm} for a realistic primordial kaon azimuthal anisotropy, $v_2 = 0.06\times\pt/$\gevc, up to $\pt =2$~\gevc, above which a constant $v_2$ is taken.
The average multiplicity density of primordial kaons have been measured previously~\cite{STAR:2008med}. In this study we use $dN_{K^+}/dy=dN_{K^-}/dy=9.35$. The event-by-event multiplicities within $|\eta|<1$ are generated from Poisson distributions with twice these averages. For simplicity, we generate primordial kaons with a uniform  pseudorapidity distributions within $|\eta|<1$.

Particle reconstruction efficiency is usually $\pt$ dependent~\cite{STAR:2008med}. 
In our simulations, we take the following $\pt$-dependent single-kaon reconstruction efficiency, 
\begin{equation}
    \varepsilon(\pt) = -0.074 + 0.64\exp{{(-0.04/\pt^{2.19})}} + 0.16\pt^{1/3}\,,
    \label{eq:eff}
\end{equation}
and include a lower $\pt$ cut of $\pt>0.1$~\gevc.  %and $\pt < 10$~\gevc} and used as the detector effect
This is similar to the parameterization in Ref.~\cite{STAR:2008med}.

We use the $|\eta| < 1$ cut to mimic the effect of finite acceptance such as that of the STAR experiment.

We now describe the \ds\ analysis procedure treating the MC data as if they were real data.
To identify decay kaons in data we would first go through all $\kk$ pairs in an event and fill a kaon pair $\minv$ histogram for these pairs and another $\minv$ histogram for rotated pairs where the $K^-$ is rotated by $\pi$ in azimuth. Rotated pairs are used to estimate the background in the real-event $\kk$ pairs. 
For each real and rotated kaon pair, inside the $\phi$-mass window, we fill {\em single} particle 3-D histograms in $(\pt,\eta,\phi-\psi_2)$ for $K^+$ and  $K^-$, separately. 
The decay kaon kinematic 3-D histograms are the difference of the two, after proper normalization of the rotated kaon pairs, for example, by the side-band method matching the real pair and rotated pair $\minv$ distributions just outside the $\phi$ mass region. 
For illustration, we can think of 1-D $\pt$ distributions. The $\pt$ projection of the real pair 3-D $(\pt,\eta,\phi-\psi_2)$ histogram minus that of the properly normalized rotated pair 3-D histogram is approximately the decay kaon $\pt$ distribution.

Additionally, we fill the 3-D histograms for all single kaons, without forming pairs of $K^+$ and  $K^-$ (i.e., looping over all kaons in the single-particle loop, not within the pair double loop). 
The reason to do this is because we will want to weight all measured kaons to match the decay kaon kinematics since the decay kaons are not identified exclusively, but only statistically.  
%We take the ratio of the 3-D histograms of decay kaons over all kaon, and then apply this ratio as a weight to all kaons in data.
We divide the ``decay kaon'' 3-D histograms by the all-single-kaon 3-D $(\pt,\eta,\phi-\psi_2)$ histogram; for $K^+$ and $K^-$, respectively. The result is a single-particle weight histogram for $K^+$ (or $K^-$) that depends on the kaon's $(\pt,\eta,\phi-\psi_2)$. To apply this weight in data, we take each kaon, find its corresponding kinematic bin in the weight histogram and assign the bin content as a weight. In this way, we are using real data kaons and mimicking them as $\phi$-decay kaons in terms of the full kinematics. Lastly, we loop through all the measured $K^+$ and $K^-$ in each event and fill the rotated and mixed-event pair $\Drho$ histograms as a function of $\ptpair$ with these weights applied at the track level. In real data analysis, the mixed events are done within the same centrality bin and with proximate primary vertex positions.

The above procedure would be used in a real data analysis. For the illustration  of  a data-driven correction procedure, we will also simply use the MC tagged $\phi$-decay kaons to obtain the weight histograms.

\begin{figure}[hbt]
    \includegraphics[width=\linewidth]{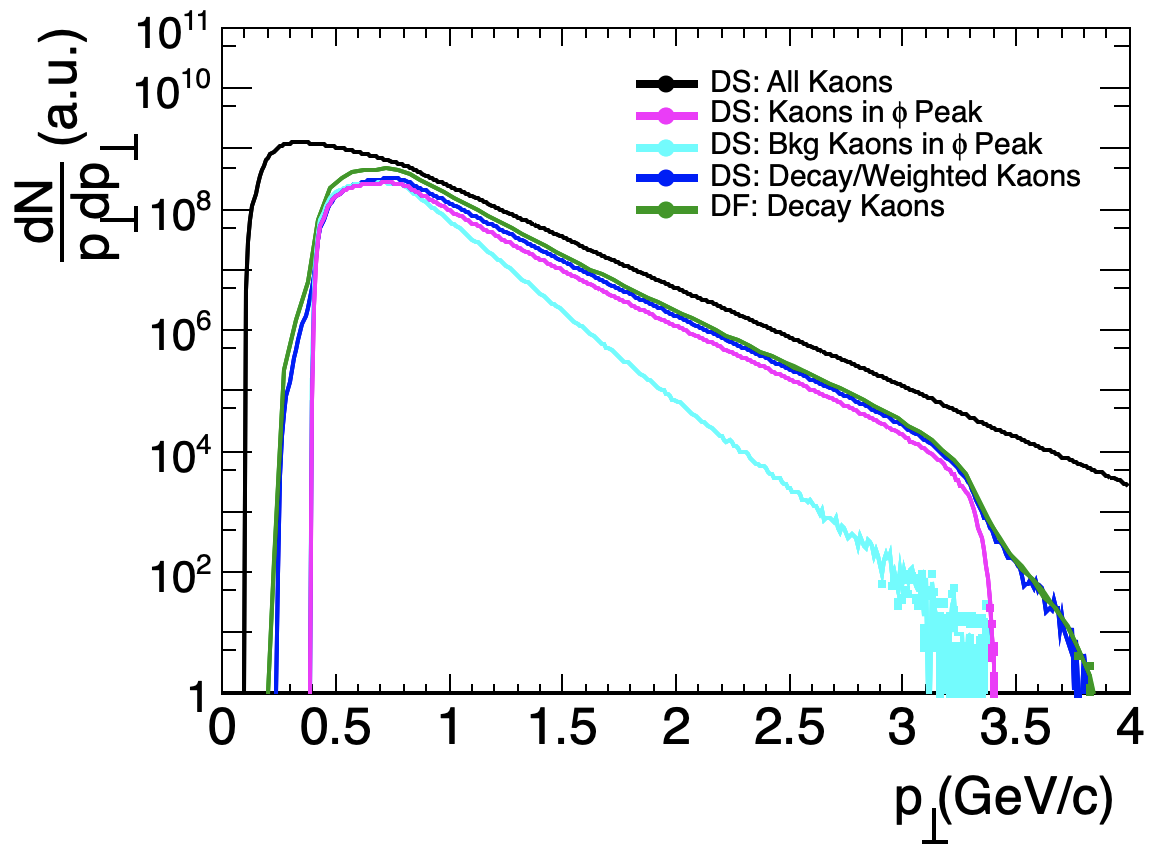}\hfill
    \caption{(Color online) Single kaon $\pt$ spectra of various sources. Black: all kaons (primordial kaons and $\phi$-decay kaons); Blue: all kaons weighted to match decay kaon kinematics; Magenta: kaons from $\kk$ pairs inside the $\phi$ mass window containing both signal and background; Cyan: rotated $\kk$ pairs inside the $\phi$ mass window, used for combinatorial background subtraction; Green: $\phi$-meson decays only from \df\ (DF). All other histograms are for kaons from \ds\ (DS) which have suffered detector effects. All histograms are arbitrarily normalized to only compare the shapes. 
    The $\phi$-mesons are generated with a uniform distribution in $|\eta|<1$, the measured $\pt$ distribution within $1.2<\pt<5.4$~\gevc, and the parameterized $v_2(\pt)$; the global spin alignment parameter for $\phi$-meson decays is set to $\Drho=0$.
    This figure illustrates the statistical identification procedure; we use MC tagging for the $\phi$-decay kaons in this study except otherwise noted.}
    \label{fig:PtCheck} 
\end{figure}

As a sanity check, we fill a 3-D histogram of all single  $K^+$ and  $K^-$ with these weights applied to see the effect of the kinematic weighting. 
Figure~\ref{fig:PtCheck} shows the effect of the weights on the single-kaon $\pt$ histogram (all histograms are arbitrarily normalized to just compare the shapes). 
All kaons are shown in black, and the blue histogram shows all kaons after weighting to match decay kaon kinematics. The green histogram shows the spectrum of $\phi$-decay kaons using published $\phi$-meson data. It is slightly softer than the reconstructed $\phi$-decay kaon spectrum (i.e., effectively the blue histogram)  because of the $\pt$-dependent efficiency applied to our MC simulation data. The magenta and cyan histograms show the intermediate $\pt$ spectra of measured kaons and rotated kaons from pairs falling within the $\phi$-meson $\minv$ peak region of [1.015,1.025]~\gevcc. They are softer than the inclusive kaons (black histogram) because of the $\minv$ requirement. 
The difference between the magenta and cyan (after side band scaling) histograms is the decay-kaon $\pt$ spectrum, effectively the blue histogram (arbitrarily normalized).

At this point, we have identified the decay kaon kinematics in ``data,'' scaled the ``measured'' single kaon kinematics to match the decay kaons, and formed rotated kaon pairs and mixed-event kaon pairs scaled to the decay-kaon pairs in kinematics. 
The difference between the $\Drho$\ of data pseudo-$\phi$ pairs from \ds\ in this subsection and that of pseudo-$\phi$ pairs from \df\ in the previous subsection is the effect of imperfect detector and finite acceptance.

\begin{figure}[hbt]
    \includegraphics[width=0.95\linewidth]{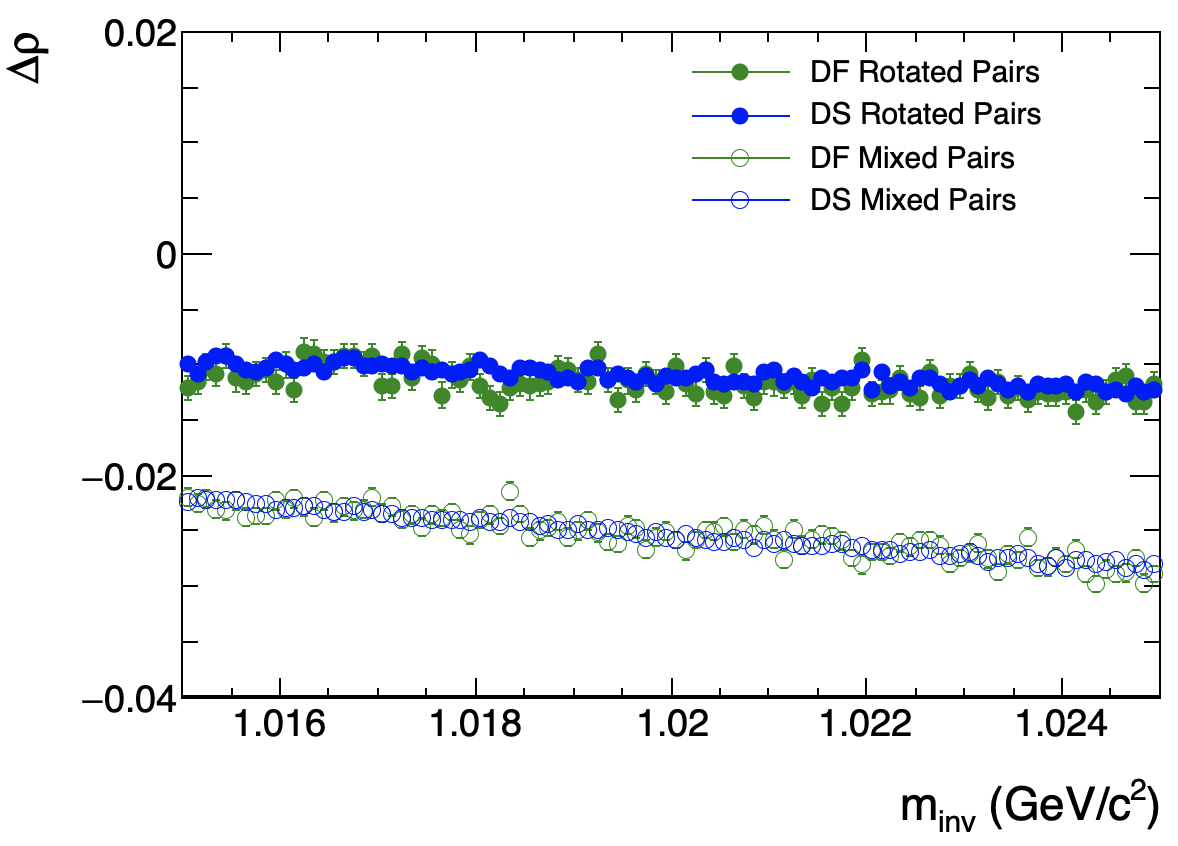}\hfill
    \caption{(Color online) $\Drho$ vs $\minv$ of pseudo-$\phi$ pairs from \df\ (DF, green points) and data pseudo-$\phi$ pairs from \ds\ (DS, blue points). 
    Filled points correspond to rotated pairs and open points correspond to mixed-event pairs. 
    The $\phi$ mesons are generated with a uniform distribution in $|\eta|<1$, the measured $\pt$ distribution within $1.2<\pt<5.4$~\gevc, and the parameterized $v_2(\pt)$; the global spin alignment parameter for $\phi$-meson decays is set to $\Drho=0$.}
    \label{fig:DataDriveDrhoMass} 
\end{figure}

In Fig.~\ref{fig:DataDriveDrhoMass}, we compare the $\Drho$\ vs  $\minv$ from \df\ and \ds. 
The mixed-event pair $\Drho$\ is smaller than the rotated pair $\Drho$. 
This is because $\Drho$\ depends on the EP resolution; In the case of mixed-event pairs, the EPs between the two events are not the same and this can be considered as zero EP resolution. EP resolution is described in~\cite{Poskanzer:1998yz}, and does affect $\phi$-meson spin alignment measurements~\cite{STAR:2022fan}; however, the EP resolution correction is straightforward and can be separated from corrections for  detector effects we focus on in this study. 
The difference between \df\ (green data points) and \ds\ (blue data points) is the $\Drho$ corrections for detector effects, respectively, for rotated and mixed-event pseudo-$\phi$ pairs. %i.e., in this simulation study, the effect of the single kaon $\pt$ efficiency. 
Note, while the input $\phi$ mesons are generated within $|\eta|<1$ and $1.2<\pt<5.4$~\gevc, the pseudo-$\phi$ kinematics are smeared and we do not cut on the kinematics of these pseudo-$\phi$ pairs in either \df\ or \ds; the average over all pseudo-$\phi$ pairs is the $\Drho$ correction.
 
\begin{figure}[hbt]
    \xincludegraphics[width=0.9\linewidth,pos=nwlow,label=\hspace*{2mm}a)]{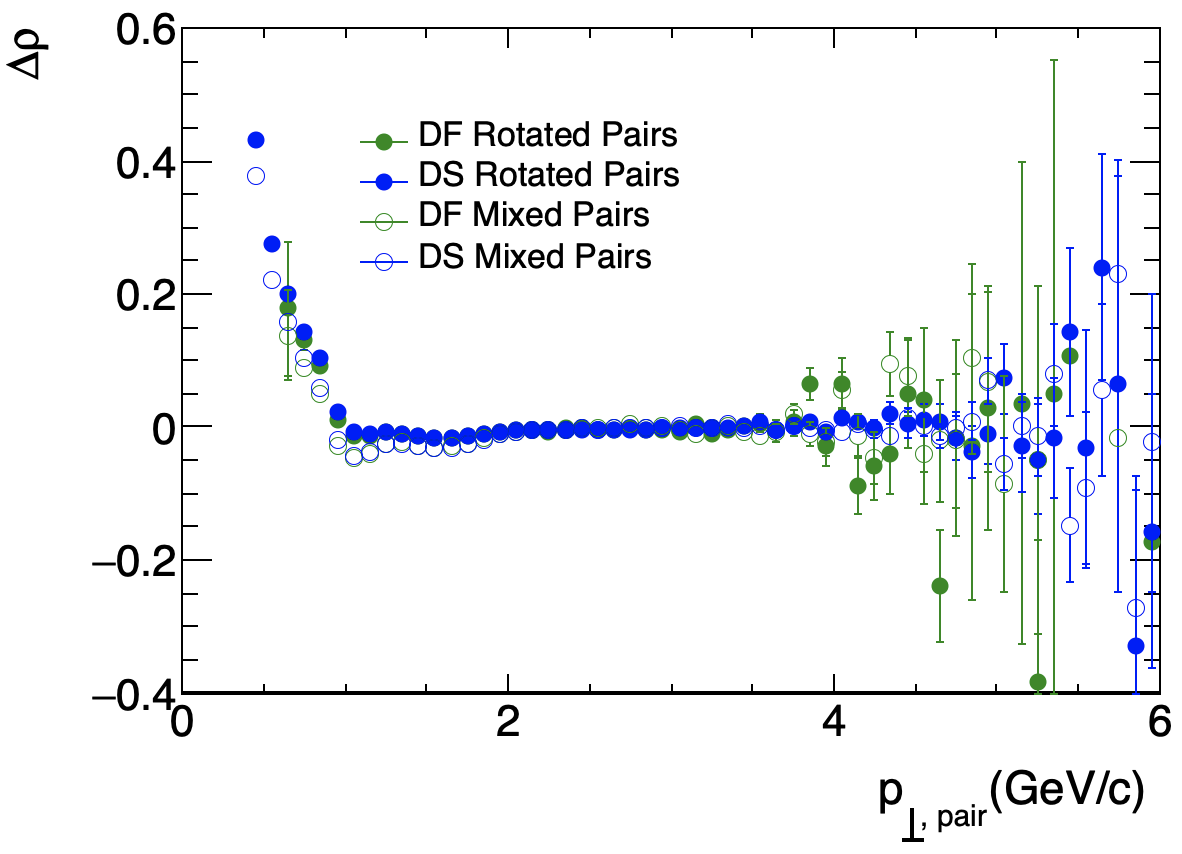}%\hfill

    \xincludegraphics[width=0.9\linewidth,pos=nwlow,label=\hspace*{2mm}b)]{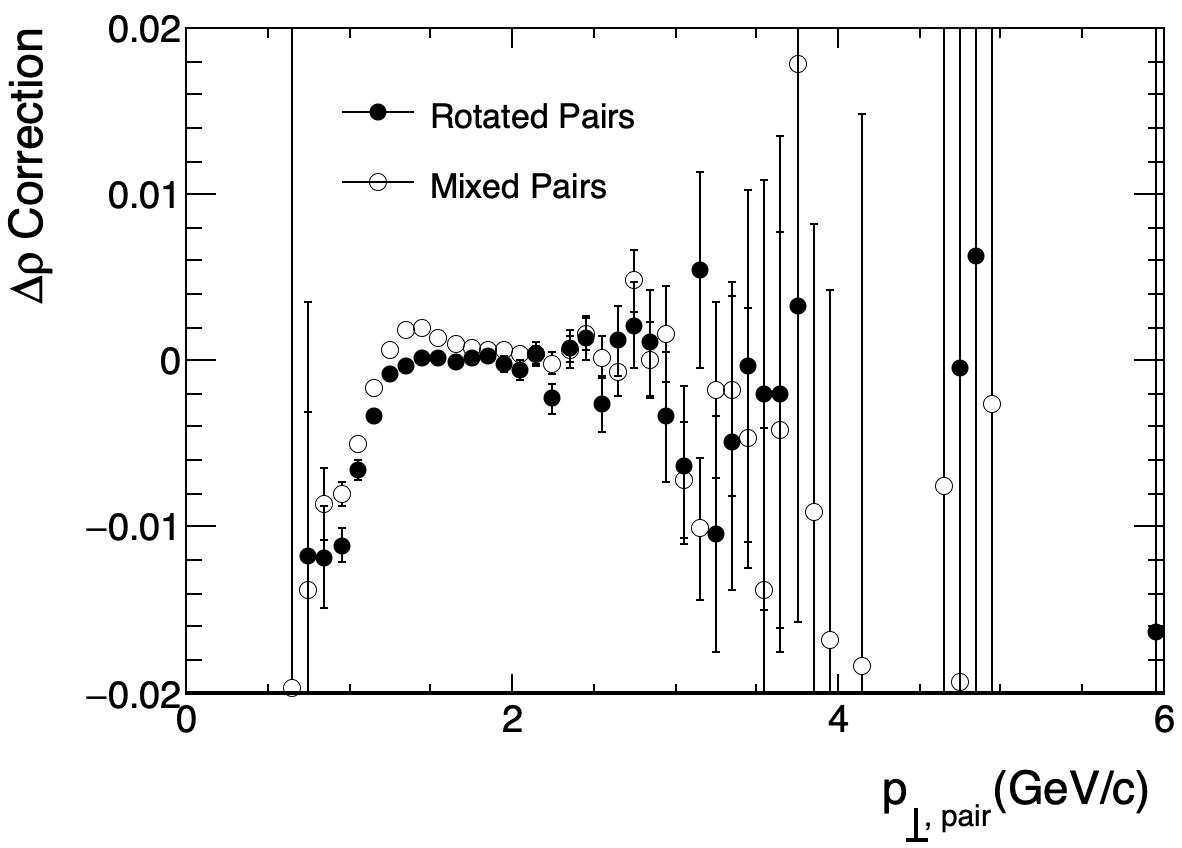}%\hfill
    \caption{(Color online) (a) $\Drho$ vs $\ptpair$ of pseudo-$\phi$ pairs from \df\ (DF, green points) and data pseudo-$\phi$ pairs from \ds\ (DS, blue points).  
    Filled markers correspond to rotated pairs and open markers to mixed-event pairs. 
    (b) Differences in $\Drho$ between green and blue markers in (a) for rotated and mixed-event pairs, respectively. We take the average of the difference (rotated pairs preferred than mixed-event pairs) across all $\ptpair$ as the detector effect on the real $\phi$ mesons.
    The $\phi$ mesons are generated with a uniform distribution in $|\eta|<1$, the measured $\pt$ distribution within $1.2<\pt<5.4$~\gevc, and the parameterized $v_2(\pt)$; the global spin alignment parameter for $\phi$-meson decays is set to $\Drho=0$.
    }
    \label{fig:ScaleExample}
\end{figure}

Figure~\ref{fig:ScaleExample}(a) shows the $\Drho$, obtained from $\mean{\cos^2\theta^*}$ via Eq.~(\ref{eq:conv}), as a function of $\ptpair$ for rotated and mixed-event pseudo-$\phi$ pairs in filled and open points, respectively. 
The \df\ pseudo-$\phi$ pairs (i.e., before any detector effects) are shown in green and the \ds\ pseudo-$\phi$ pairs (i.e., after detector effects) are shown in blue. Figure~\ref{fig:ScaleExample}(b) shows the difference in $\Drho$ between \df\ and \ds, which is the detector correction from our data-driven approach. The corrections are relatively small, and are somewhat different ($\sim 0.002$) between using rotated pairs and mixed-event pairs. This difference arises from EP mismatch in the mixed-events, so the rotated pair correction is more trustworthy.

%================================================================================================
\section{MC Closure Test}\label{sec:closure}
We examine the closure of our data-driven method using toy-model MC. 
The MC generation of $\phi$-mesons, their decays, and implementation of the detector acceptance/efficiency effects to the decay daughter kaons (as well as the generated primordial kaons) are described in Sections~\ref{sec:folding} and~\ref{sec:scaling}.
The MC truths of the detector effects for (i) the real $\phi$ kaon pairs and (ii) the pseudo-$\phi$ kaon pairs are precisely known, by examining those that have survived detector effects of finite acceptance and track reconstruction inefficiency of single kaons (referred to as survived real and pseudo $\phi$ pairs, respectively).
We take the pseudo-$\phi$ kaon pairs as a proxy for the real $\phi$-decay kaon pairs to study the detector effects.
The comparison between (i) and (ii) tells us how valid this approach  (i.e.~{\em Assumption I} in Section~\ref{sec:DataDrivenMethod}) is.
We obtain (iii) the detector effect in our data-driven method from \ds\ and \df. 
We can compare the detector effect from (iii) to the MC truths of (i) and (ii) to check whether our data-driven procedure can faithfully represent them. In an ideal case, all three should be the same.

\begin{figure}[hbt]
    \xincludegraphics[width=0.9\linewidth,pos=swHigh,label=\hspace*{3mm}a)]{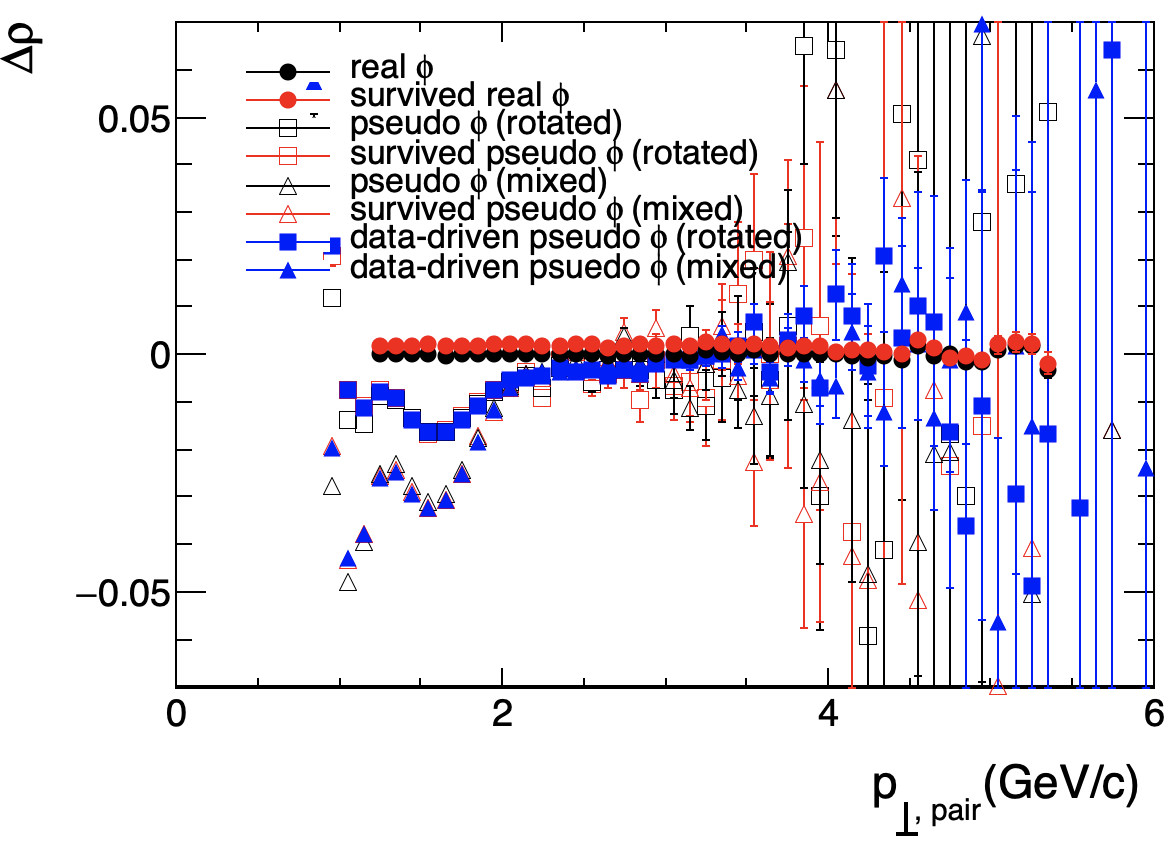}%\hfill
    
    \xincludegraphics[width=0.9\linewidth,pos=swHigh,label=\hspace*{2mm}b)]{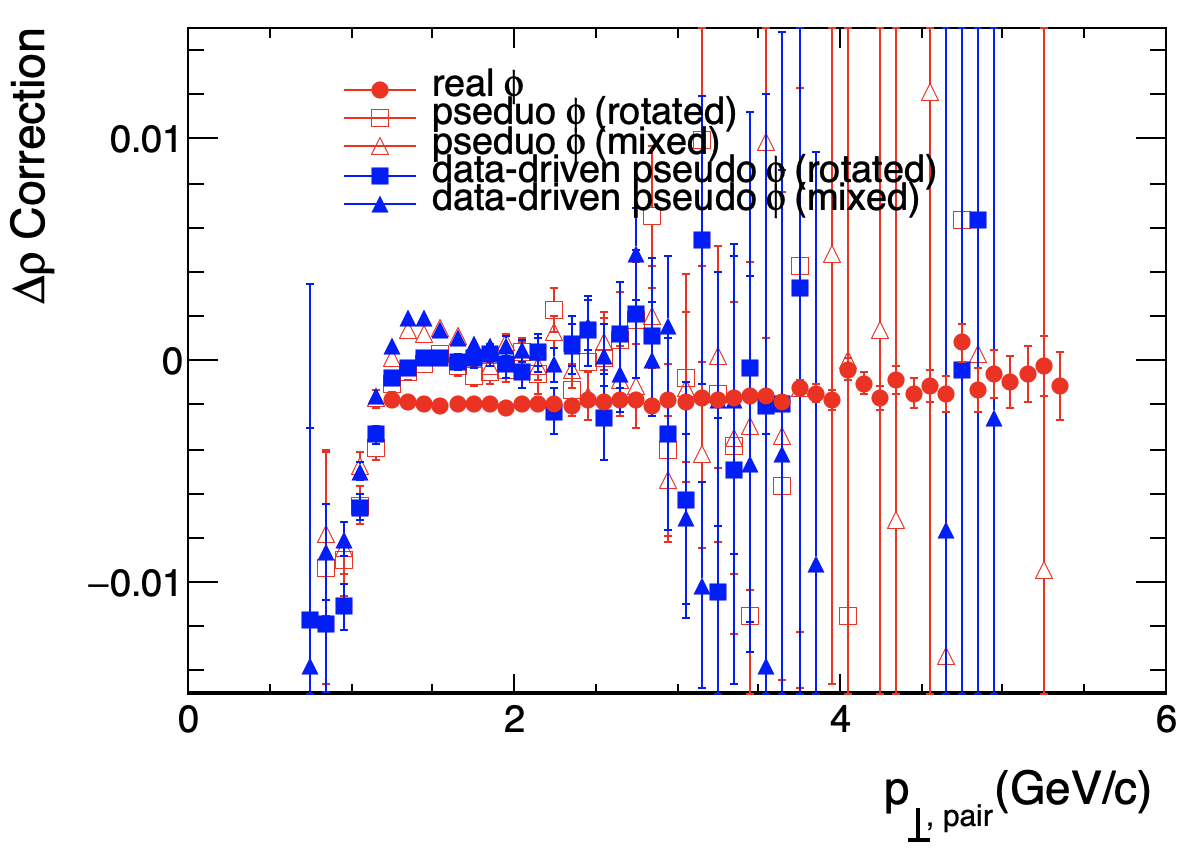}%\hfill
    \caption{(Color online) (a) $\Drho$ vs $\ptpair$ for real-$\phi$ pairs (black circles) and pseudo-$\phi$ (black squares/triangles) pairs from MC input and, correspondingly, for the survived real-$\phi$ pairs (red circles) and survived pseudo-$\phi$ (red squares/triangles) pairs, and for data pseudo-$\phi$ (blue squares/triangles) pairs from \ds\ where all single kaons (primordial and decay) are weighted to have the same single-particle kinematics as those from $\phi$-meson decays. (b) The red points show the differences of the black points minus the corresponding red points in (a), representing the truth $\Drho$ corrections for detector effects for real $\phi$ (solid red points) and pseudo $\phi$ (open red points), respectively. The blue squares/triangles show the differences of the black squares/triangles (\df) minus the blue squares/triangles (\ds) in panel (a); These are the $\Drho$ corrections derived from the data-driven method using rotated/mixed-event pseudo-$\phi$ pairs. The $\phi$ mesons are generated with a uniform distribution in $|\eta|<1$, the measured $\pt$ distribution within $1.2<\pt<5.4$~\gevc, and the parameterized $v_2(\pt)$; the global spin alignment parameter for $\phi$-meson decays is set to $\Drho=0$.}
    \label{fig:CombPhiThe}
\end{figure}

Figure~\ref{fig:CombPhiThe}(a) shows the $\Drho$'s of the generated real $\phi$ (black filled circles), rotated pseudo-$\phi$ (black open squares/triangles) and, correspondingly in red, those of survived real $\phi$ and survived pseudo $\phi$. 
The $\Drho$ is zero for the generated $\phi$ mesons, as expected, because $\Drho=0$ was assigned to the decay of those $\phi$ mesons. The detector effects are seen to create a positive $\Drho\sim 0.003$ for the survived real $\phi$'s [the red filled circles in Fig.~\ref{fig:CombPhiThe}(a)] approximately independent of the $\phi$-meson $\pt$. The $\pt$ dependencies of $\Drho$\ of the real $\phi$ and pseudo $\phi$ are different, and this is presumably because of the different pair kinematics between the two. 
The detector effects do not seem to cause much a change in $\Drho$ of the pseudo $\phi$, as seen from comparing the $\Drho$'s of pseudo $\phi$ (black open squares/triangles) and survived pseudo $\phi$ (red open squares/triangles).These effects can be more clearly shown in Fig.~\ref{fig:CombPhiThe}(b) by the differences before and after detector effects in the solid circles and open squares for real $\phi$ and pseudo $\phi$, respectively. There exists a difference in the detector effects between the real $\phi$ and pseudo-$\phi$ shown in Fig.~\ref{fig:CombPhiThe}(b); The difference is on the order of $\sim0.002$.

We now perform the ``data-driven'' correction method using \df\ and \ds. \df\ is simply the open black squares/triangles in Fig.~\ref{fig:CombPhiThe}(a), i.e.~the pseudo-$\phi$ kaon pairs without any detector effects. 

\ds\ weights all (decay and primordial) kaons to have the same single-particle kinematics as the decay kaons, and then forms the data pseudo-$\phi$ pair $\Drho$ of those kaons with detector effects. 
This weighting can be done by statistically identifying decay kaons as in real data analysis or by tagging tracks in MC simulation. We use the latter in this MC closure study because our purpose is to see how well the detector effects can be corrected, not about how well the $\phi$-meson signal is reconstructed by combinatorial background subtraction; we checked that the difference between using these two ways is small (see below). The $\Drho$ of data pseudo-$\phi$ pairs obtained from \ds\ is shown in Fig.~\ref{fig:CombPhiThe}(a) in blue squares/triangles. 
Similarly, the difference between \df\ (black squares/triangles) and \ds\ (blue squares/triangles) in Fig.~\ref{fig:CombPhiThe}(a) is shown in Fig.~\ref{fig:CombPhiThe}(b). The results show that the data-driven procedure captures the detector effects of those MC-tagged pseudo-$\phi$ pairs (shown in red open squares/triangles). 

\begin{figure}[hbt]
    \includegraphics[width=0.95\linewidth] {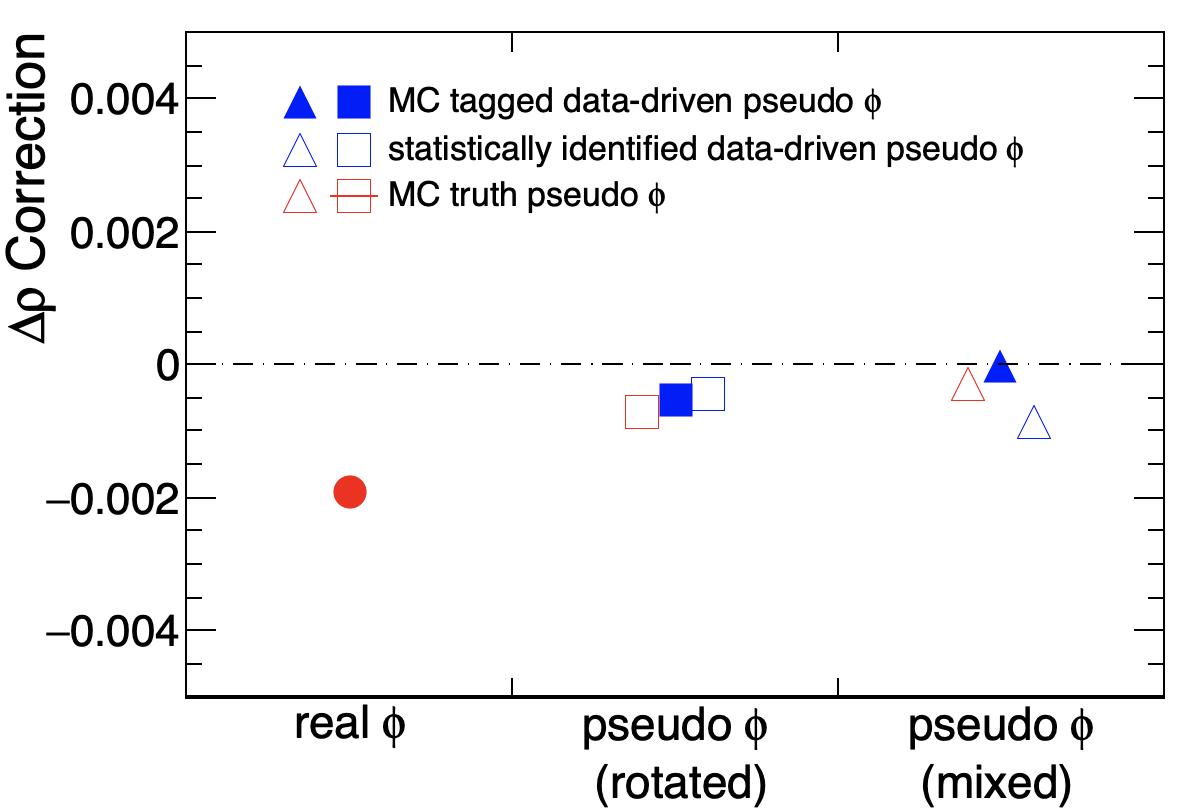}\hfill
    \caption{(Color online) Average $\Drho$ detector correction for real $\phi$ (solid circle), pseudo $\phi$ from rotated pairs (squares), pseudo $\phi$ from mixed-event pairs (triangles). Pseudo $\phi$ includes the MC pseudo $\phi$ (red square/triangle) and data pseudo $\phi$ from MC Tagging (filled blue square/triangle) and from statistically identifying kaons (open blue square/triangle).
    The $\phi$ mesons are generated with a uniform distribution in $|\eta|<1$, the measured $\pt$ distribution within $1.2<\pt<5.4$~\gevc, and the parameterized $v_2(\pt)$; the global spin alignment parameter for $\phi$-meson decays is set to $\Drho=0$.
    }
    \label{fig:DetEffComp}
\end{figure}

Figure~\ref{fig:DetEffComp} shows the detector effect corrections for real $\phi$, pseudo-$\phi$ rotated and mixed-event pairs, and data-driven corrections from \df\ $-$ \ds\ using rotated and mixed-event pairs. 
In the case of mixed events, there is an EP mis-match, which can cause discrepancy in the observed detector correction. In Fig.~\ref{fig:DetEffComp}, the filled blue points correspond to using MC-tagging information to weight the kaons, and the open blue points correspond to statistically identified kaons for the purpose of weighting. The difference is relatively small between these two ways. As seen from Fig.~\ref{fig:DetEffComp} comparing the blue points to their corresponding red points, 
the data-driven corrections capture the true detector effects for the pseudo $\phi$, 
validating {\em Assumption II} of our data-driven correction procedure (see Section~\ref{sec:DataDrivenMethod}).
However, there exists a discrepancy in the corrections between real $\phi$ (red filled circle) and pseudo $\phi$ (red square/triangle) shown in Fig.~\ref{fig:DetEffComp}, as previously seen in Fig.~\ref{fig:CombPhiThe}. As noted previously, the discrepancy (MC non-closure) is on the order of 0.002, 
indicating a breakdown of {\em Assumption I} or an accuracy of the assumption of $\sim 0.002$. This is comparable to the experimental systematic uncertainty of the $\Drho$ measurements~\cite{STAR:2022fan}.
%--------------------------------------------------------------------------------------------------
\section{Decipher the $\Drho$ Correction}\label{sec:DrhoDependence}

\subsection{Dependence on $\phi$-meson $\Drho$ signal}
\begin{figure}[hbt]
    \includegraphics[width=0.95\linewidth] {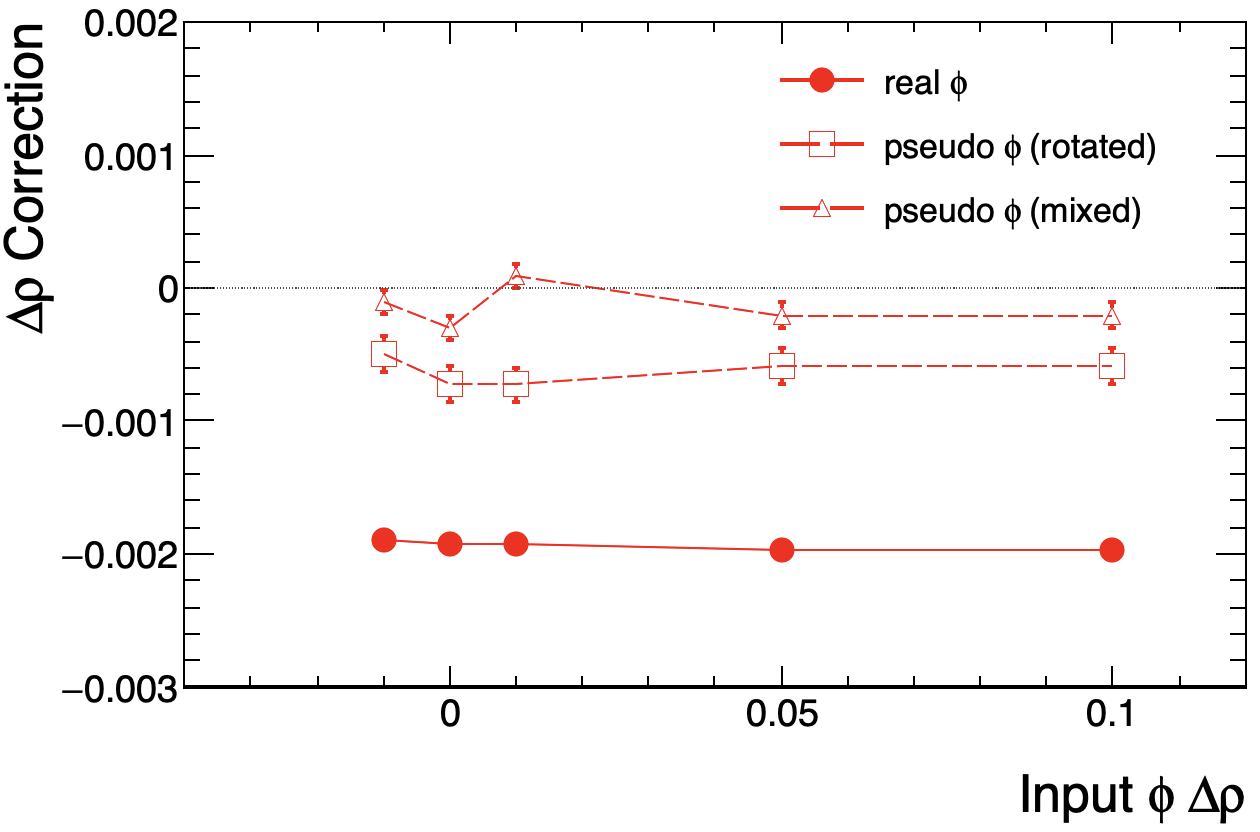}\hfill
    \caption{(Color online) $\Drho$ detector corrections for real $\phi$ (solid circles), pseudo $\phi$ from rotated pairs (open squares), and pseudo $\phi$ from mixed-event pairs (open triangles) as a function of the input $\phi$-meson $\Drho$. 
    The $\phi$ mesons are generated with a uniform distribution in $|\eta|<1$, the measured $\pt$ distribution within $1.2<\pt<5.4$~\gevc, and the parameterized $v_2(\pt)$.
    The detector corrections are largely independent of the input $\phi$-meson $\Drho$. 
    }
    \label{fig:DrhoDependence}
\end{figure}

In real data, the $\phi$-meson $\Drho$ may not be zero~\cite{ALICE:2019aid,STAR:2022fan}. However, the detector correction (obtained through embedding or data-driven) should ideally not depend on the input $\phi$-meson $\Drho$ signal. 
We check this by varying the input $\phi$-meson $\Drho$ and obtaining the detector correction for real $\phi$ and pseudo $\phi$. This check is shown in Fig.~\ref{fig:DrhoDependence}, where for reasonable input values of $\Drho$ the correction on the real $\phi$ is found indeed to be independent of input signal. The correction for the pseudo $\phi$ is also independent of the input signal, and this is easy to understand because the pseudo-$\phi$ pairs do not contain real $\phi$-decay pairs and therefore would not know about the input $\Drho$ signal. 
A finite difference exists between the corrections for real $\phi$ and pseudo $\phi$ as seen previously in Section~\ref{sec:closure}. 

\subsection{Dependence on $\phi$-meson $\pt$}
\begin{figure*}[hbt]
\centering
    \xincludegraphics[width=0.333\textwidth,pos=nw,label=a)]{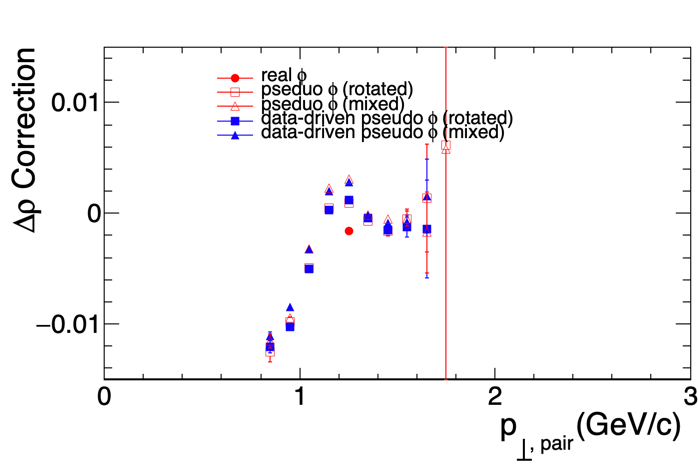}\hfill
    \xincludegraphics[width=0.333\textwidth,pos=nw,label=b)]{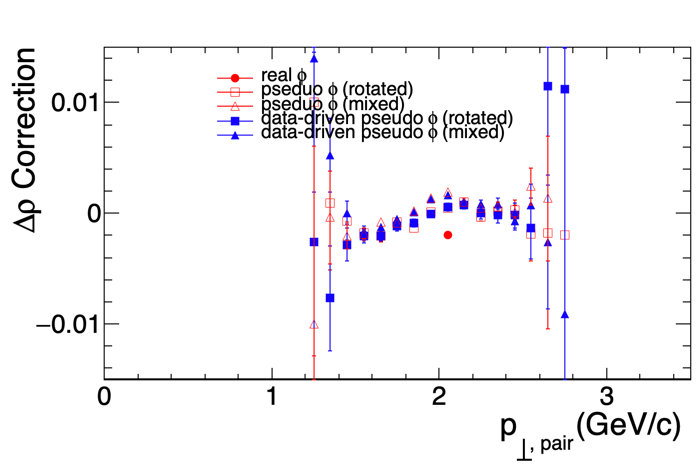}\hfill
    \xincludegraphics[width=0.333\textwidth,pos=nw,label=c)]{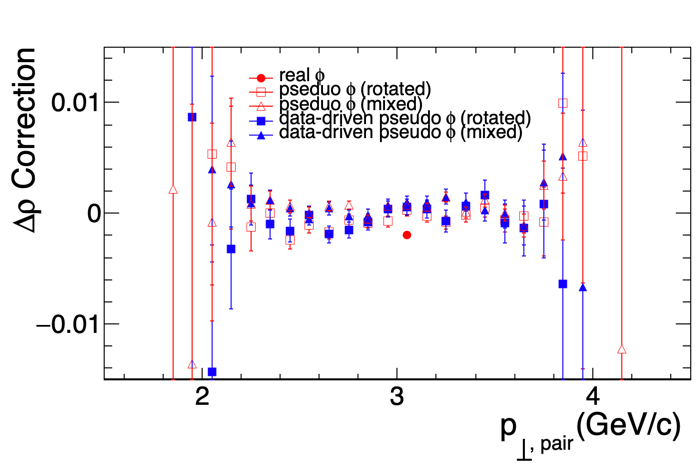}\hfill
    \caption{(Color Online) As same as Fig.~\ref{fig:CombPhiThe}(b) but for fixed $\phi$-meson input $\pt$ of (a) 1.2~\gevc, (b) 2~\gevc, and (c) 3~\gevc. 
    The $\phi$ mesons are generated with a uniform distribution in $|\eta|<1$ and the parameterized $v_2$ at the given $\pt$ value; the global spin alignment parameter for $\phi$-meson decays is set to $\Drho=0$.
    The data points are non-monotonic as a function of $\ptpair$ and generally appear to peak at the fixed value of the input $\phi$-meson $\pt$ (for the two lower input $\pt$ values).}
    \label{fig:PhiSinglePtCorPt}

\end{figure*}

We have used so far a finite range of $1.2<\pt<5.4$~\gevc\ for input $\phi$ mesons. 
To gain more insights into the non-closure in $\Drho$ of $\sim 0.002$ between real $\phi$ and pseudo $\phi$, we repeat the analysis using fixed $\phi$-meson $\pt$ values of 1.2, 2, and 3~\gevc.
Figure~\ref{fig:PhiSinglePtCorPt} shows the corresponding results similar to Fig.~\ref{fig:CombPhiThe}(b). The correction for the real $\phi$ is a single point at the fixed $\pt$ value.
The $\ptpair$ of the pseudo-$\phi$ kaon pair is not single-valued but smeared. 
Interestingly, these $\Drho$ corrections are not monotonic in $\ptpair$, but peaked at the fixed value of the input $\phi$-meson $\pt$. 
One possible way to understand this (and the non-monotonicity over $\ptpair$) may be as follows. The pseudo-$\phi$'s at the input $\pt$ value are most alike the real-$\phi$'s, and those further away in $\ptpair$ will deviate from the peak $\Drho$ value in the same, apparently decreasing direction. Because of the decreasing trend, it is natural to expect the pseudo-$\phi$ $\Drho$ value at the input $\pt$ value to be larger than the input real-$\phi$ $\Drho$, so that the average $\Drho$ of the pseudo-$\phi$'s is closer to the input real-$\phi$ $\Drho$ value. The average of the $\Drho$ corrections over all these pseudo $\phi$'s would be our $\Drho$ correction.
The idea behind this correction procedure is that the average over $\ptpair$ for both the real-$\phi$ and pseudo-$\phi$ pairs would ideally be the same. 

\begin{figure}[hbt]
    \includegraphics[width=0.95\linewidth]{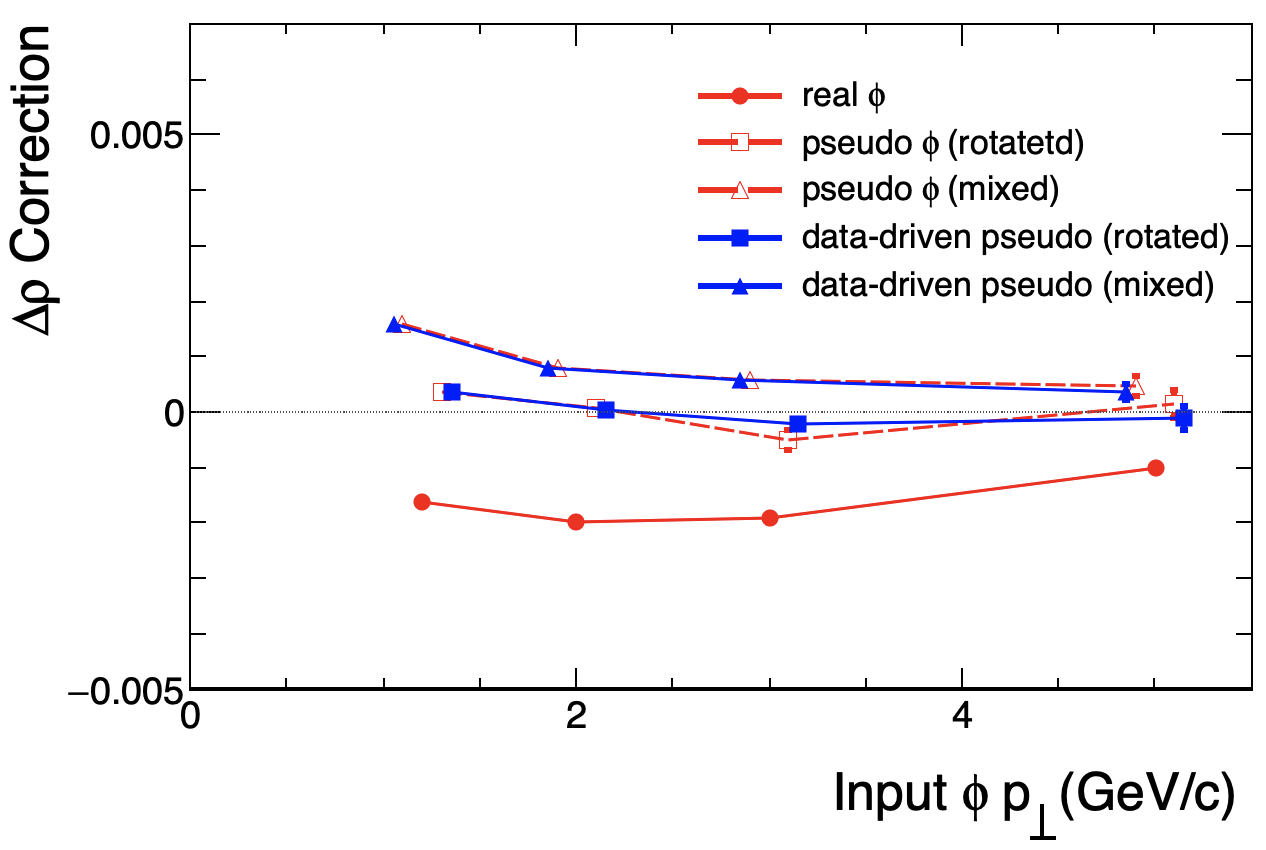}\hfill
    \caption{ (Color Online) $\Drho$ detector corrections for real $\phi$ of various fixed $\pt$ values (filled red circles), for pseudo $\phi$ from rotated/mixed-event pairs (red squares/triangles), and the corresponding data-driven corrections for pseudo-$\phi$ using rotated/mixed-event pairs (blue squares/triangles), points shifted for clarity. The $\phi$ mesons are generated with a uniform distribution in $|\eta|<1$ and the parameterized $v_2$ at the given $\pt$ values; the global spin alignment parameter for $\phi$-meson decays is set to $\Drho=0$.}
    \label{fig:CorrectionPtDep}
\end{figure}

Figure~\ref{fig:CorrectionPtDep} shows the $\Drho$ corrections as a function of the input $\phi$-meson $\pt$ of the fixed-$\pt$ study. The correction for pseudo $\phi$ differs from that for real $\phi$ by approximately 0.002 as previously observed; The difference is somewhat more significant at lower $\pt$. 
Again, the data-driven correction from the rotated pseudo-$\phi$ pairs generally reproduce the truth correction for pseudo $\phi$; the performance using mixed-event pairs is poorer.

\subsection{Dependence on $\phi$-meson $v_2$}
The detector correction depends on the $\phi$-meson kinematics such as the $\phi$-meson $\pt$ and may thus also depend on the $\phi$-meson elliptic flow $v_2$. In this subsection, we investigate the corrections for the real $\phi$, psuedo $\phi$, and our data driven correction (using rotated or mixed events) as a function of the input $\phi$-meson  $v_2$. 

\begin{figure}[hbt]
    \includegraphics[width=0.95\linewidth]{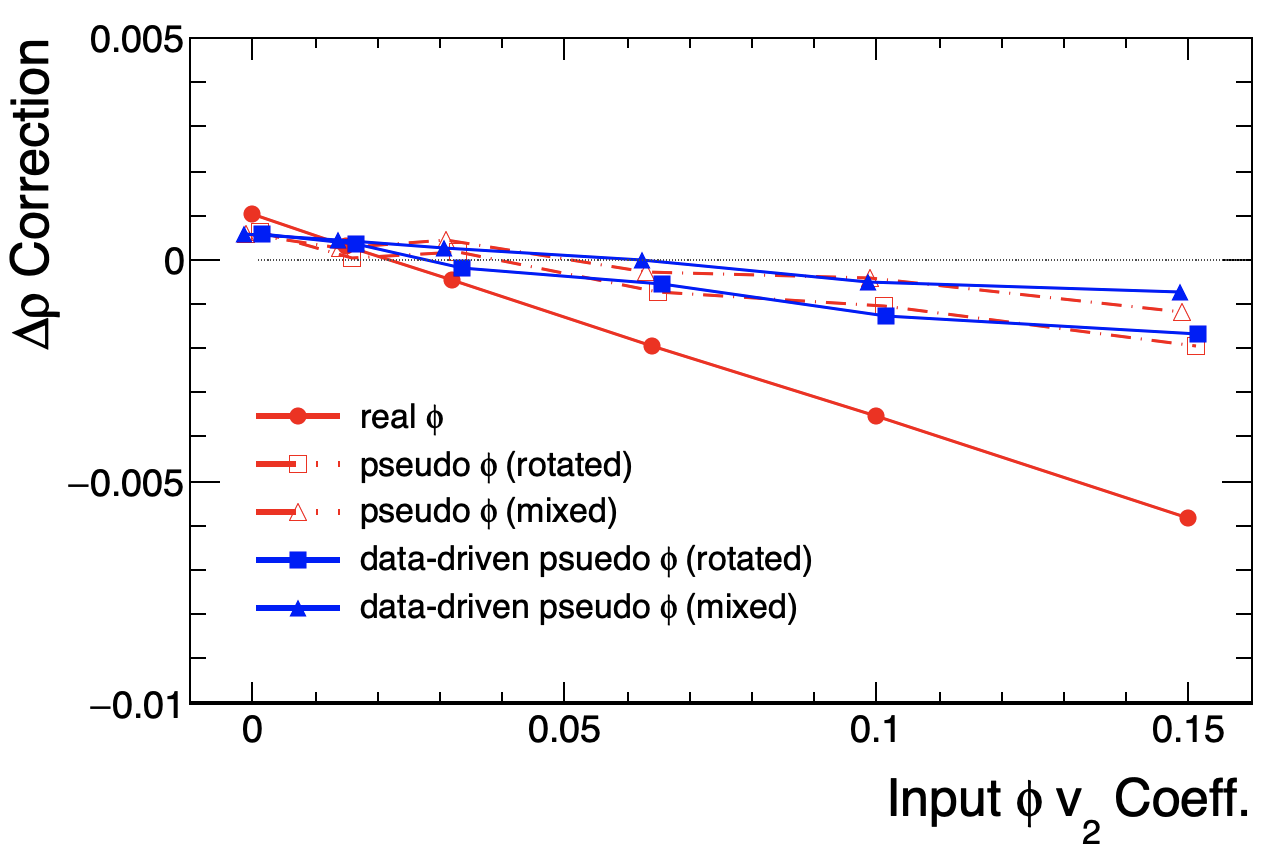}\hfill
    \caption{(Color online) $\Drho$ detector corrections for real $\phi$ (red circles), pseudo $\phi$ from rotated/mixed-event pairs (red squares/triangles), and the data-driven corrections for pseudo $\phi$ using rotated/mixed-event pairs (blue squares/triangles) for different input $\phi$-meson $v_2(p_{\perp})$ coefficients, points shifted for clarity. The $\phi$ mesons are generated with a uniform distribution in $|\eta|<1$ and the measured $\pt$ distribution within $1.2<\pt<5.4$~\gevc; the global spin alignment parameter for $\phi$-meson decays is set to $\Drho=0$.}
    \label{fig:V2Dependence}
\end{figure}

The $\phi$-meson $v_2$ is taken to be linear in $\pt$, 
\begin{equation}
    v_2=\hat{v}_2\times\pt/\mbox{\gevc}\,,
\end{equation}
up to $\pt=3$~\gevc\ above which the $v_2$ is set to be constant as done in Section~\ref{sec:folding}.
. 
Figure~\ref{fig:V2Dependence} shows $\Drho$ detector correction as a function of the $v_2$ coefficient, $\hat{v}_2$. The detector correction for real $\phi$ decreases nearly linearly with $v_2$ towards more negative values. This can be understood to arise from the coupling of $v_2$ and decay orientation-dependent $\kk$ pair reconstruction efficiency, as follows. A decay along the $\phi$-meson momentum direction will result in a $\kk$ pair with asymmetric momenta in the lab frame and, because of the $\pt$-dependent efficiency, such a pair is more likely lost (the extreme case is that one of the daughter kaons has too low a momentum that it is always lost). On the other hand, a decay perpendicular to the $\phi$-meson momentum direction will yield a $\kk$ pair with equal momentum magnitude in the lab frame, thus more likely to be reconstructed. With finite $v_2$ (i.e., more $\phi$ mesons in the RP direction), there will be more kaon pairs reconstructed perpendicular to the RP resulting in a positive $\Drho$ (or a negative correction). This effect is proportional to $v_2$, and this is approximately shown by the real-$\phi$ correction result in Fig.~\ref{fig:V2Dependence}.
The $v_2$ dependence is much weaker for the pseudo $\phi$. This is because the pseudo $\phi$ is formed effectively from pairs of kaons from decays of two different $\phi$'s so the $v_2$ effect is smeared out (the effect is likely second order in $v_2$ and thus negligible). There is a noticeable $v_2$ dependence for pseudo $\phi$ in Fig.~\ref{fig:V2Dependence}, and we will come back to this point.
 
It is interesting that the $\Drho$ correction starts at a positive value at zero input $v_2$. With $\phi$-meson $v_2 =0$, there can still be detector effects from the $\pt$-dependent efficiency, even though the efficiency is symmetric in the transverse plane (independent of azimuth). This is because $\theta^*$ is a three-dimensional angle and the $\pt$-dependent efficiency distorts the decay $\theta^*$ distribution in the rest frame, so there can be complicated effects even when the $\phi$-meson $v_2 =0$.

In addition, with $\phi$-meson $v_2=0$, we still expect some finite detector acceptance effect as well. The acceptance effect can be understood by the $\eta$ cut distorting the $\phi$-decay $\theta^*$ distribution in the pair rest frame away from a spherical one. This acceptance effect is from the longitudinal direction and is therefore finite even when the $\phi$-meson $v_2 =0$.
The acceptance effect is generally $\pt$ dependent--the lower the $\pt$ the more significant effect from an acceptance $\eta$ cut. This may cause an effect on $\Drho$ by the coupling between the finite acceptance and the $\pt$-dependent efficiency.The $\Drho$ measurement depends generally on the kaon kinematics.  
As the $\phi$-meson $v_2$ affects the $\pt$ distribution of the decay kaons and causes the $\pt$ distribution to vary with azimuthal angle, the acceptance effect can be complex and can have a dependence on the $\phi$-meson $v_2$. All these subtle effects may be responsible for the small but noticeable $v_2$ dependence of the pseudo $\phi$ $\Drho$ correction seen in Fig.~\ref{fig:V2Dependence}.
These effects should also be present for the real-$\phi$ $\Drho$ correction, but may be subdominant to the significant ``direct'' $v_2$ dependence.

As shown in Fig.~\ref{fig:V2Dependence}, the difference between real-$\phi$ and pseudo-$\phi$ $\Drho$ corrections increases with $v_2$ because of the dominant $v_2$ dependence in the former and the less sensitivity of the latter to the $\phi$-meson $v_2$. The finite $v_2$ breaks {\em Assumption I} of our data-driven method, whereas at small $v_2$ {\em Assumption I} is approximately fulfilled.
In all cases, the data-driven approach reproduces the pseudo-$\phi$ $\Drho$ correction, fulfilling {\em Assumption II}.

\section{Summary}
In this work we have developed a data-driven method to correct for detector effects on $\phi$-meson global spin alignment measurements. Data-driven corrections are preferred to MC or ``embedding'' corrections because the detector effects are inherited in the data and thus precise, whereas MC simulations are never perfect. The data-driven method uses pseudo-$\phi$ kaon pairs, formed from rotated pairs of decay kaons or from mixed events within the $\phi$-meson mass window, as a proxy for real $\phi$ mesons. 
The main motivation to use pseudo-$\phi$ kaon pairs is that they are expected to experience the same detector effects as the real-$\phi$ decay kaon pairs and not expected to have any physical spin alignment except those caused by kinematics. The assumption here ({\em Assumption I}) is that the detector effects on $\Drho$ are equal for pseudo-$\phi$ and real-$\phi$.

Because the $\phi$ mesons are not exclusively identified in relativistic heavy ion collisions, we use the \ds\ technique to mimic those pseudo-$\phi$ kaon pairs, described in Section~\ref{sec:scaling}. The \ds\ forms pseudo-$\phi$'s from all measured kaons in data by weighting them to have the same single-particle kinematics as those $\phi$-decay kaons. To avoid the real-$\phi$ signals in the data, one charge-sign of the kaons (e.g., $K^-$) is rotated by $\pi$ in azimuth, or the kaon pair is taken from mixed events.
The $\phi$-decay kaon kinematic distributions can be obtained statistically in a real data analysis. 
The $\Drho$ of these data pseudo $\phi$'s from the \ds\ (statistical) technique represents the $\Drho$ of the pseudo $\phi$'s that have survived detector effects of finite acceptance and reconstruction inefficiency. 
The assumption here ({\em Assumption II}) is that the detector effects on $\Drho$ are equal for the data pseudo $\phi$ from \ds\ and the pseudo $\phi$ from \df.

In order to get the detector correction to $\Drho$, the $\Drho$ value before any detector effects is needed. 
This is obtained by the \df\ technique, described in Section~\ref{sec:folding}, generating $\phi$ mesons according to the measured kinematic distributions and decaying them with a given $\Drho$. 
One then form pseudo-$\phi$ pairs from the decay kaons (either rotated or from mixed events) and calculate their $\Drho$.
The difference in $\Drho$ between the \df\ pseudo $\phi$'s and \ds\ data pseudo $\phi$'s is the correction for detector effects. 

We perform a MC closure test of this data-driven method. 
It is found that {\em Assumption II} is fulfilled, namely, the $\Drho$ correction for data pseudo $\phi$'s from \ds\ appears to always equal that for pseudo $\phi$'s from \df, using rotated pairs. 
The correction for pseudo $\phi$'s is found to be generally small, on the level of $-0.001$.
It is also found, however, that {\em Assumption I} is not fulfilled, namely, the $\Drho$ correction for pseudo $\phi$ does not {\em always} equal that for real $\phi$. 
The discrepancy is found to be caused mainly by elliptic flow $v_2$ of the $\phi$ mesons and is approximately proportional to $v_2$ (see Fig.~\ref{fig:V2Dependence}). 
This $v_2$ dependence arises from the decay orientation-dependent pair reconstruction efficiency coupled with $v_2$, resulting in a positive $\Drho$ (or negative $\Drho$ correction) for the real $\phi$ mesons.
The $v_2$ effect is largely smeared for pseudo $\phi$ because they are formed from combinatorial pairs of kaons effectively from decays of different $\phi$ mesons or from mixed events. 
Residual $v_2$ dependence seems to remain in the pseudo $\phi$ $\Drho$ correction, presumably because of the intertwining nature of kaon kinematics, acceptance and efficiency, as well as the boost to the pair rest frame; for example, the $v_2$ alters the decay kaon kinematics as a function of the azimuthal angle causing an acceptance effect that may depend on $v_2$. 
The disparity in the responses to $v_2$ between real $\phi$ and pseudo $\phi$ results in the discrepancy, or a breakdown of {\em Assumption I}.
 
For the $\phi$-meson $v_2$ magnitude measured in relativistic heavy-ion collisions, the $\Drho$ corrections appear to be negative for both pseudo $\phi$ and real $\phi$, approximately $-0.0005$ and $-0.002$, respectively. The discrepancy is approximately 0.001--0.002, and is within the typical experimental systematic uncertainties for $\phi$-meson global spin alignment measurements. However, additional studies and comparisons with standard experimental MC correction procedures may be needed as the global spin alignment signal is small, possibly less than $1\%$.

\section*{Acknowledgment} 
This work is supported in part by the U.S.~Department of Energy (Grant No.~DE-SC0012910).

\bibliography{ref}
\end{document}